\documentclass[prb,twocolumn,  preprintnumbers,amsmath,amssymb]{revtex4-1}
\usepackage{graphicx}
\usepackage{dcolumn}
\usepackage{bm}
\usepackage{subfigure}
\usepackage{multirow}
\usepackage{amsmath}
\usepackage{color,soul}
\usepackage[font=footnotesize, justification=raggedright,singlelinecheck=false]{caption}
\begin{document}

\title{Self-Assembled Chiral Photonic Crystals From  Colloidal Helices Racemate}
\author{Qun-li Lei$^{1,2}$, Ran Ni$^{2, \dag}$ and Yu-qiang Ma$^{1 \dag}$}

 \affiliation{
   $^1$ National Laboratory of Solid State Microstructures and Department of Physics, Collaborative Innovation Center of Advanced Microstructures,\\
          Nanjing University, Nanjing 210093, China\\
   $^2$    School of Chemical and Biomedical Engineering,\\
          Nanyang Technological University, 637459, Singapore\\
  $\dag$ E-mail: r.ni@ntu.edu.sg (R. N.); myqiang@nju.edu.cn (Y. M.) }

\begin{abstract}

{ \textbf{Abstract:}}
Chiral crystals consisting of micro-helices have many optical properties while presently available fabrication processes limit their large-scale applications in photonic devices. Here, by using a {simplified} simulation method, we investigate a bottom-up self-assembly route to build up helical crystals from the  {smectic} monolayer of colloidal helices racemate. With increasing the density, the system undergoes an entropy-driven co-crystallization by forming crystals of various symmetries with different helical shapes. In particular, we identify two crystals of helices arranged in the binary honeycomb and square lattices, which are essentially composed by two sets of opposite-handed chiral crystal. Photonic calculations show that these chiral structures can have large complete photonic bandgaps. In addition, in the self-assembled chiral square crystal, we also find dual polarization bandgaps that selectively forbid the propagation of circularly polarized lights of a specific handedness along the helical axis direction. The self-assembly process in our proposed system is robust, suggesting possibilities of using chiral colloids to assemble {photonic metamaterials}.
\end{abstract}

\maketitle

Photonic crystals with complete photonic bandgaps (PBGs) provide a versatile platform  for optical communications,\cite{russell2003} information technology,\cite{dolev2010} solar energy harvesting,\cite{Wehrspohn2012} and medical diagnostics,\cite{Fenzl2014} \textit{etc.}.  Although a few structures, \textit{e.g.}, diamond and gyroid crystals,\cite{maldovan2004d,dolan2015optical,turner2013,saba2011} are known to have outstanding photonic properties, the fabrication of large-size photonic crystals with 3D PBGs in the visible region is still highly challenging.\cite{hynninen2007self,dolan2015optical,liu2016diamond,ducrot2017colloidal,
long2018,cersonsky2018}
Recently, it was demonstrated that some  crystals formed by ordered micro helical particles can have full omnidirectional PBGs.\cite{chutinan1998s,toader2001p} The intrinsic twist and chirality of helices can also produce other interesting optic phenomena,  like polarization PBG,\cite{thiel2007p,lee2005pol,science2009goldhelix,
esposito2014n,kao2015d,chen2008polarization}
circular polarization conversion,\cite{kaschke2015} negative refraction\cite{optic2010PRL} and coreless fibre wave guidance.\cite{wong2012excitation,beravat2016t,russell2017helically} Variation of the lattice symmetry, the direction or the relative phase of neighbouring helices in the lattice can change its optic behaviour significantly.\cite{thiel2010t,thiel2009t} For example, photonic topological insulators with robust topological protected states have been realized in honeycomb helical arrays.\cite{rechtsman2013p} Several anomalous topological phases\cite{leykam2016edge,leykam2016a} including photonic Weyl point\cite{noh2017experimental} were found in the square helical array where adjacent helices have $180^{\circ}$ phase shift.
These exotic topological phases have shown the promise of using helical particles/fiber to fabricate  photonic metamaterials with unconventional properties.\cite{kaschke2016,alexeyev2013spin,kartashov2013dynamics,
alexeyev2016localized,petrovic2017rotating,lu2016topological}

Helical or spiral crystals in visible/IR spectrum were first fabricated by glancing angle deposition techniques,\cite{kennedy2002f} in which the deposited helices have the same phase. In principle, the focused {electron}/ion beam induced deposition,\cite{esposito2014n} the direct laser writing\cite{thiel2007p} and the holographic lithography\cite{pang2005c,raub2011large} can build up helical crystals with arbitrary symmetry, while the slow process of these techniques sets a bottleneck to the large-scale fabrication.  Compared with top-down building of helical crystal whose complexity scales up with the number of helices, fabrication of free helical colloids or fiber with controllable chirality is more technically accessible using current nano-techniques,\cite{schamel2014,mark2013hybrid,tottori2012magnetic,
zhang2009artificial,yuan2018chiral} chirality-transferring chemical synthesis methods,\cite{grason2015chirality,JACS2012transfer,LiuMinghua2015chiral,
LiuMinghua2015,colloid2003JACS,feng2017assembly} or even biological helical structures like DNA tubes,\cite{maier2017self} spiral bacteria,\cite{young2006selective,adler2006genus} bacterial flagella,\cite{Dogic2006PRL} {spirulina algae\cite{yan2017multifunctional} }and plant vessel.\cite{gao2013}
Although there are several theoretical and simulation studies trying to understand the self-assembly  mechanism of helical particles{\cite{Ferrarini2013isotropic,Frezza2014left,cinacchi2017,cinacchi2016,Ferrarini_kolli2016,
Ferrarini2014self,Dijkstra2014,dussi2015,theory2010generic}} and fiber,\cite{grason2015Rev,hall2016}  no attempt has been made to connect the self-assembled structures with their optical properties.

\begin{figure*}[t]
\centering
  \includegraphics[height=7.0cm]{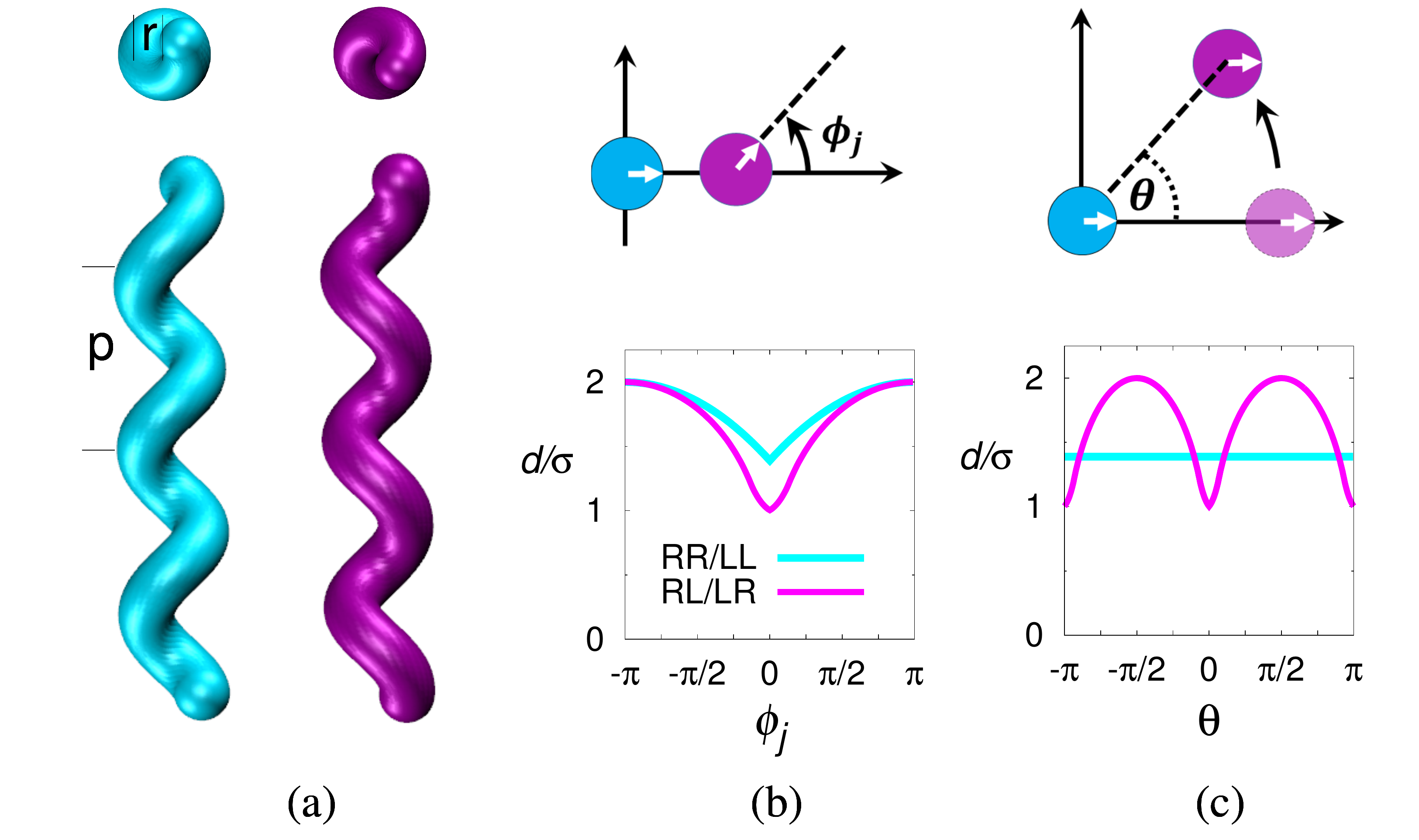}
  \caption{ \label{fig1} (a) Top and side views of two parallel helices with opposite chirality. (b) Pairwise chiral interactions represented by the contact distance $d_{ij}$ ($i,j$=R,L) as a function of the self-rotating angle $\phi_j$ of helix $j$. The arrows on spheres indicate the self-rotating angles. The reference helix $i$ is at the origin and its self-rotating angle is $\phi_i=0$. (c) $d_{ij}$ as a function of the angle of position $\theta$ of helix $j$ with respect to helix $i$. Here $[r,p]=[0.5 \sigma, 3.0\sigma]$. For both (b), (c), $z_j = z_i$. }
\end{figure*}

In this work, we propose a simulation method, in which the computation of interaction between long parallel helices can be much simplified by their screw-axis symmetry. With this technique, we investigate a bottom-up self-assembly route to build up chiral helical crystals based on the entropy-driven co-crystallization of opposite-handed colloidal hard helices. We find a variety of self-assembled crystals with different helical shapes. Self-supported helical square lattices with $180^{\circ}$ nearest-neighbour phase-shift and hexagonal lattice with $\pm 120^{\circ} $ nearest-neighbour phase-shift {are} robustly assembled, and both crystals have large complete PBGs. With further optimization, the minimal dielectric contrast for the square helical crystals to have PBGs can be as low as 3.5, surpassing the diamond crystals of dielectric spheres.\cite{joannopoulos2011p} More interestingly, in the self-assembled square helical crystals, we also find dual polarization PBGs for circularly polarized lights along the axes of helices, which selectively allows one type of circularly polarized light to propagate through. This suggests the possibilities of designing self-assembling {colloidal helices} for large-scale fabrication of chiral beamsplitters with the application as logic gates in photonic quantum computing.\cite{turner2013,o2009photonic,lodahl2017}

\section*{Results}
\subsection*{Pairwise Interaction Between Parallel Hard Helices}
\begin{figure*}[t]
\centering
		\resizebox{190mm}{!}{\includegraphics[trim=0.0in 0.0in 0.0in 0.0in]{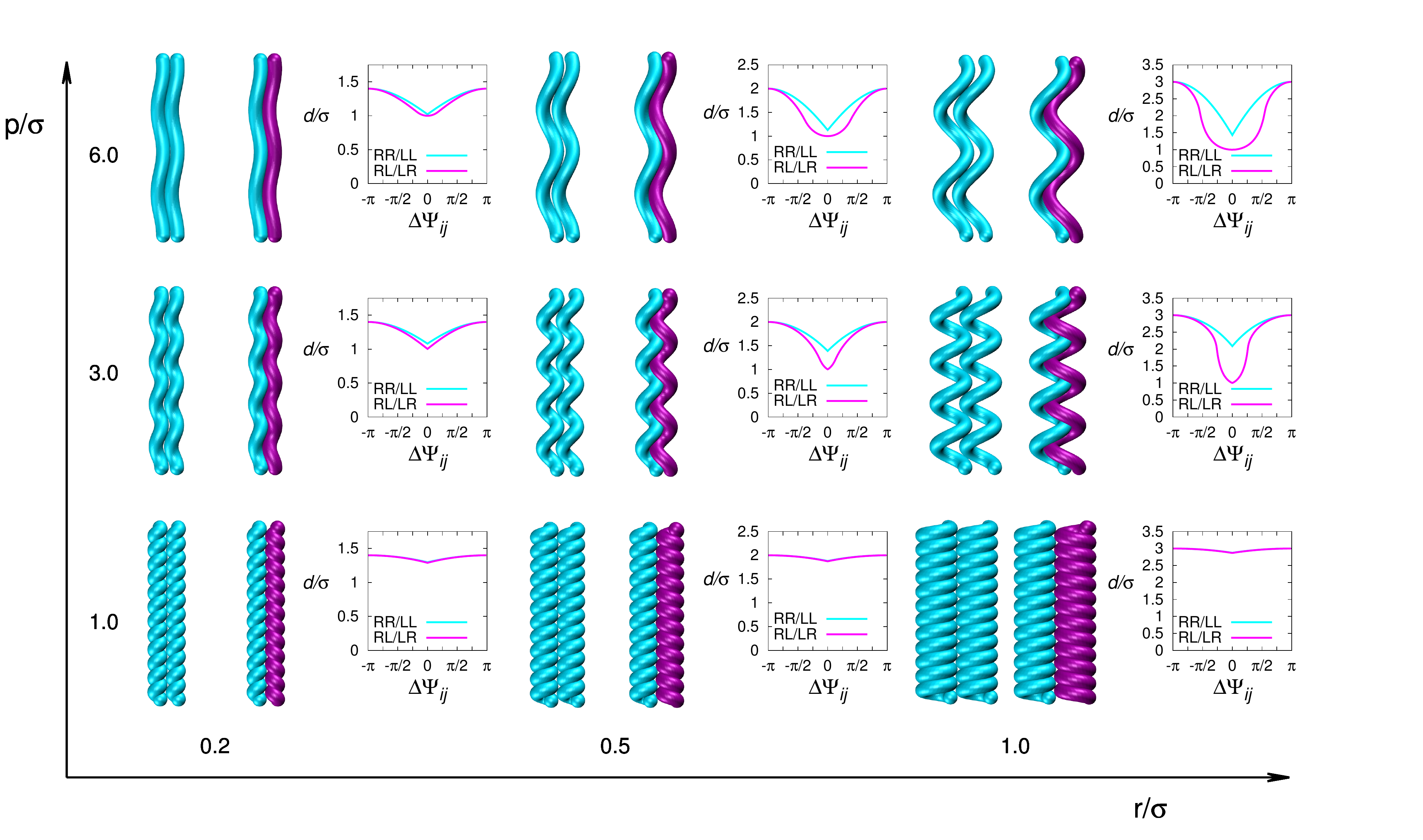} } 
\caption{\label{interaction_phase} Pairwise chiral interactions between two parallel helices of arbitrary handedness with different geometric parameter $r$ and $p$. The diagram for each configuration shows the contact distance $d_{ij}$ as a function of the interaction phase $\Delta \Psi_{ij}$. }
\end{figure*}

We model colloidal helices as twisted rods whose cross-section is a disk of diameter $\sigma$, and the shape is controlled by the helical radius $r$ and helical pitch $p$ (Figure \ref{fig1}a). We assume the orientations of all helices are the same. Due to the chiral shape of helix, the excluded volume interaction between two parallel like-chiral helices is different from that of the opposite-chiral pair. This lateral excluded volume interaction  can be characterized by the pairwise contact distance  $d_{ij}$, \textit{i.e.}, the allowed minimal distance between the central axes of two parallel helices without overlap. The subscript $i$ refers to  the reference helix, and the subscript $j$ refers to  the approaching helix. $i,j=R, L$ as both helices can be right-handed (R) or left-handed (L) helix.  Due to the mirror symmetry,  $d_{RR}=d_{LL}$ and $d_{RL}=d_{LR}$. As a result of the anisotropic shape of helices, $d_{ij}$ depends not only on the self-rotating angle of helices $\phi_{i/j}$, the elevation of the helices in the axial direction $z_{i/j}$, but also on the relative angle of the position of helix $j$ with respect to helix $i$ which is denoted as $\theta$. In Figure \ref{fig1}b, we plot $d_{ij}$ as a function of  $\phi_j$  by fixing $\phi_i=0$, $\theta=0$, and $z_{i} = z_j$ for typical helices with $[r,p]=[0.5 \sigma,3.0\sigma]$. We find periodic oscillation of both $d_{RR/LL}(\phi_j)$ and $d_{RL/RL}(\phi_j)$  with period of $2\pi$.  The dependence of $d_{ij}$ on $\theta$ with $\phi_i=\phi_j=0$ and $z_j = z_i$ is shown in Figure~\ref{fig1}c. One can see that $d_{RR/LL}(\theta)$ is independent with $\theta$, while $d_{RL/LR}(\theta)$ has the periodic dependence on $\theta$ with the period of $\pi$. The behavior of $d_{ij}(z_{i/j})$ is similar to $d_{ij}(\phi_{i/j})$ thus is not shown. { 
Because of such complex anisotropic interactions between the helices, general analytical overlap-checking algorithms between two helices are currently unavailable, and helices can only be modeled as connected hard spheres.\cite{cinacchi2016,Ferrarini_kolli2016,
Ferrarini2014self,Dijkstra2014,dussi2015}} The computational cost thus scales up dramatically with the increase of coarse-grained degree and the length of helices. In Method section, we rigorously prove that {for the specific case of parallel helices}, as a result of screw axis symmetry of helices, $d_{ij}$ can be written as a periodic function of a single variable $\Delta \Psi_{ij}$, which is defined as the \emph{interaction phase}:
\begin{eqnarray}
d_{ij}(\phi_{i},\phi_{j},\theta, z_{i},z_{j} )= d_{ij}\left (\Delta \Psi_{ij}\right), \\
\Delta \Psi_{ij} = C_j\phi_{j}-C_i\phi_{i}+(C_j-C_i)\theta + 2 \pi (z_{j}-z_{i})/p, \label{interaction_phase}
\end{eqnarray}
where $C_{i/j}$ is the chirality sign of the helices\ $i/j$ ($C_R=1$, $C_L=-1$). The computation of the interaction between helices thus can be much simplified {and independent of the coarse-grained degree of helices}, if $d_{ij}(\Delta \Psi_{ij})$ is known {\em a priori}. In Figure~\ref{interaction_phase}, we show the typical behavior of $d_{ij}(\Delta \Psi_{ij})$ between two parallel helices with different geometric parameter $r$ and $p$. One can see that both $d_{RR/LL}$ and $d_{RL/LR}$ have the minimal value at $\Delta \Psi_{ij}= 0$, while the minimum of $d_{RL/LR}$ is always smaller than that of $d_{RR/LL}$. These imply that the opposite-chiral helix pair can form more compact pairwise packing compared with the like-chiral pair. We also find as $p$ or $r$ increases, the anisotropy (depth of the valley in $d_{ij}$) of the interactions become stronger for both opposite-chiral and like-chiral pairs, while the difference between the two pairwise interactions becomes most pronounced at around $p=3\sigma$. {This suggests the packing between opposite-chiral helices pairs is most favored at this region. As can be seen later, this packing advantage over like-chiral pairs is the underlying reason to form stable racemate crystal phases around $p=3\sigma$.}

\subsection*{Self-Assembly in Colloidal Helices Racemate}

\begin{figure*}[t]
\centering
  \includegraphics[height=15.0cm]{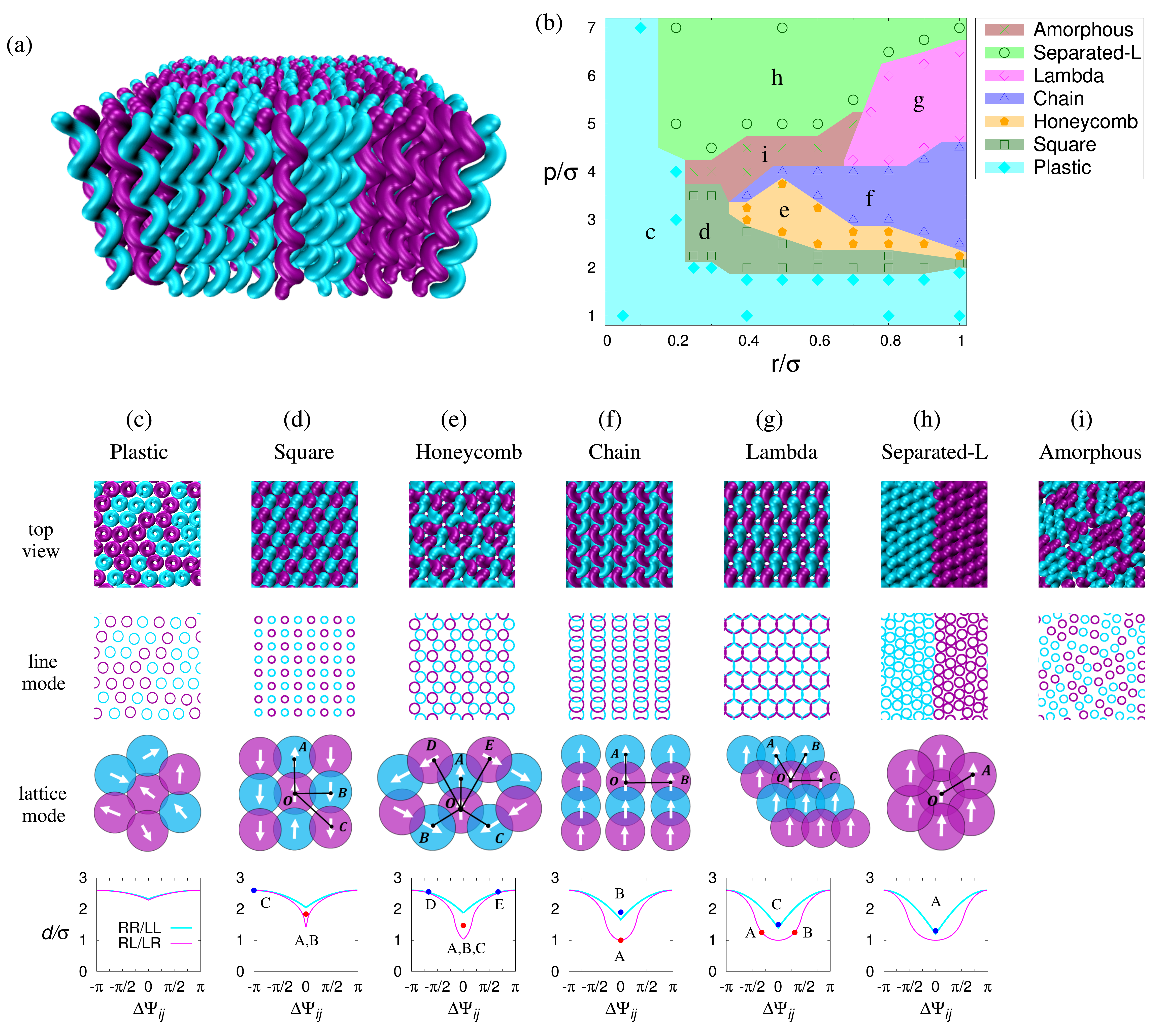}
  \caption{\label{phase_dagram}  (a) Schematic of racemic helices monolayer which contains equal numbers of right- and left-handed hard helices. (b) the self-assembly diagram in the representation of $r$ and $p$  obtained from MC simulations and  (c-i) the corresponding  self-assembled structures: (c) plastic crystal, (d) square crystal, (e) honeycomb crystal, (f) chain-like crystal, (g) lambda crystal, (h) Separated-L phase, and (i) amorphous phase. In (c-i), {The first three rows are the top views of each structure (along helical axis) with different representations. The second row shows helices in the representation of helical lines. Note that the overlap between the lines is due to the angle of view.} The third row shows the lattice schematic for each structure with arrows indicating the self-rotating orientation of each helix. In the fourth row, the contact distances and the interaction phases between the reference helix and its nearest neighbours are indicated by the solid dots on the curves of $d_{ij}(\Delta \Psi)$. The geometric parameters of the helix are $r=0.8\sigma$ and $p=1.5\sigma$(c), $2.25\sigma$(d), $2.75\sigma$(e), $3.5\sigma$(f), $5\sigma$(g), and $7\sigma$(h), and different colors indicate the helices with different chirality.}
\end{figure*}

\begin{figure*}
\centering
		\resizebox{180mm}{!}{\includegraphics[trim=0.0in 0.0in 0.0in 0.0in]{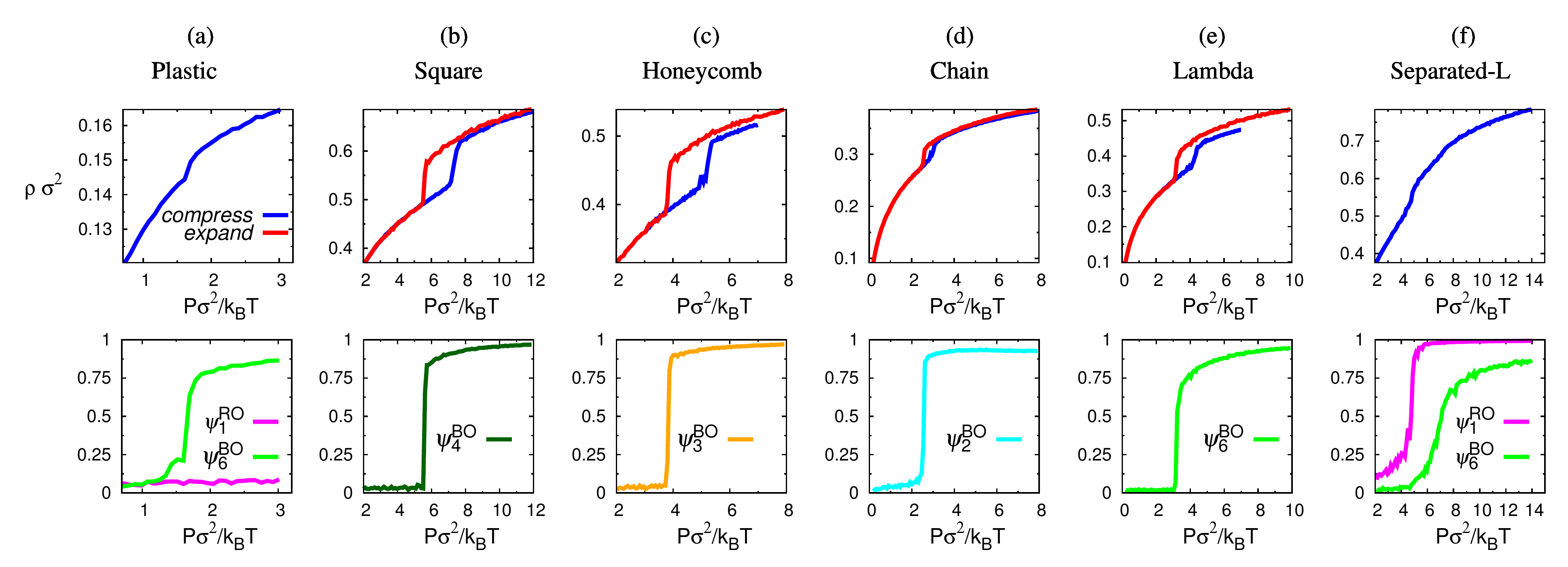} } 
\caption{\label{EOS} Equation of state (first panel), two dimensional crystalline order parameter $\psi^{BO}_i$, $\psi^{RO}_1$ (second panel)  for the formation of different structures in MC simulations. $P$ denotes the 2D pressure and $\rho$ is the 2D denisty. All the figures in second  panel correspond to the expand process except for the plastic and separated-L cases. The geometric parameters of helices for each phase are: plastic phase [$r$,$p$]=[0.8$\sigma$, 1.5$\sigma$], square crystal [$r$,$p$]=[0.3$\sigma$, 2.5$\sigma$], honeycomb crystal [$r$,$p$]=[0.5$\sigma$,3.0$\sigma$], chain crystal [$r$,$p$]=[0.8$\sigma$, 3.5$\sigma$], lambda crystal [$r$,$p$]=[0.8$\sigma$, 5$\sigma$], separated-L phase [$r$,$p$]=[0.5$\sigma$, 7$\sigma$].}
\end{figure*}

We obtain the accurate $d_{RR}(\Delta \Psi_{ij})$, and $d_{RL}(\Delta \Psi_{ij})$ (see Method), with which we perform {intensive} Monte Carlo (MC) simulations of a liquid crystal  monolayer composed by the same numbers of oppositely handed helices (racemate), as shown in Figure~\ref{phase_dagram}a. The axes of all helices are perfectly aligned. This model is the simplification of a recent experimental system,\cite{Dogic2014Nature} in which colloidal rod-like virus with weak spiral shape can form  {smectic} monolayers on the open smooth surface with polymer depletants. In our MC simulations, the initial configuration of the system is a disordered racemic fluid. We slowly increase the pressure of the system and identify the first ordered phase that nucleates. We find six different self-assembled structures with various $r$ and $p$. The self-assembly diagram in the representation of $r$ and $p$ is summarized in Figure~\ref{phase_dagram}b. {In the first three panels} of Figure~\ref{phase_dagram}c-i, we show the typical configurations of these structures from the top view (along the axis of helix)  {with different representations.} The lattice schematic and the self-rotation of helix (indicated by the arrows) in each crystal are shown in the third panel. In the fourth panel we plot $d_{ij}(\Delta\Psi_{ij})$ for each crystal and mark the interaction phases and the distances between the reference helix and its first (red) and second (blue)  {nearest} neighbours by points, \textit{e.g.}, A  B, C, \textit{etc.}, as shown in the corresponding lattice schematic. In Figure~\ref{EOS}, we also show the equation of state (EOS) for each phase in the first panel and the crystalline order parameter $\psi^{BO}_i$ and $\psi^{RO}_1$ as functions of pressure in the second panel (see Method for the detailed definition). The crystalline order parameter $\psi^{BO}_i$ indicates the formation of crystal with specific symmetry in the system. More specifically,  $\psi_6^{BO}$ is for the hexagonal lattice; $\psi_4^{BO}$ is for the square lattice; $\psi_3^{BO}$ is for the honeycomb lattice; $\psi_2^{BO}$ is for one dimension chain lattice. The self-rotating orientational order parameter $\psi^{RO}_1$, like the orientational order parameter in liquid crystal system, measures how likely that helices take the same rotating angles.

We find that when the helical pitch or radius is small ($p< 2 \sigma$ or $r<0.2 \sigma$), the self-assembled structures are racemic plastic hexagonal crystals (Figure~\ref{phase_dagram}c), in which the coordination number $n_c=6$. In these crystals, two types of helices are randomly mixed, and the system has the long-range six-fold orientational order, \textit{i.e.}, high $\psi^{BO}_6$, but without long-range correlation in the self-rotating angle, \textit{i.e.}, low $\psi^{RO}_1$ (Figure~\ref{EOS}a). The formation of this racemic plastic hexagonal crystal is due to the fact that the interaction between helices is almost  isotropic in 2D. 

As $p$ increases above $2\sigma$, the valleys of both curves $d_{RL}(\Delta\Psi_{ij})$ and $d_{RR}(\Delta\Psi_{ij})$ become deeper and the gap between the minima of $d_{RL}(0)$ and $d_{RR}(0)$ increases. Depending on $p$, the binary square ($n_c=4$) or binary honeycomb ($n_c=3$) crystal first nucleates from the fluid with increasing pressure as indicated by the increase of $\psi^{BO}_4$ and $\psi^{BO}_3$ in Figure~\ref{EOS}b,c respectively. 
The nearest neighbours in these crystals are both opposite-chiral and take the zero interaction phase $\Delta\Psi_{RL}=0$ with respect to the nearest neighbours, by which the crystal can have the most compact local packing (Figure~\ref{phase_dagram}d,e). 
On the contrary, the second-nearest helices are all like-chiral and have larger contact distances with respect to the reference helix. This leads to the fact that helices in these two crystals can touch their second-nearest neighbours before overlapping with their first-nearest neighbours with increasing the density. This excluded-volume effect between helices and their second-nearest neighbours determines the close-packed density for the square crystal  at $r>0.25\sigma$ and $p>2.1 \sigma$, and for the honeycomb crystal at $r>0.4\sigma$.  In fact, each of these two binary crystals can be viewed as a combination of two sets of interconnected chiral crystals with opposite chiralities. As shown later, these chiral crystals have 3D PBGs and the contact between like-chiral helices is important to enlarge the PBG size and mechanically support the crystals.

Further increasing $p$ makes the chain crystal ($n_c=2$) nucleate from the fluid phase for systems with the relatively large helical radius (see $\psi^{BO}_2$ in Figure~\ref{EOS}d). In these structures, the helices are first arranged in chains with an alternate handedness sequence {interlocked but without overlap with each other}, and the chains are packed side by side (Figure~\ref{phase_dagram}f). The interaction phases between helices and their first and second-nearest neighbours in this structure are both at the optimal packing position, \textit{i.e.}, $\Delta\Psi_{ij}=0$. At even larger $p$, the system nucleates into a lambda crystal (see $\psi^{BO}_6$ in Figure~\ref{EOS}e), where stripes of like-chiral helices are arranged in the alternate chirality sequence   (Figure~\ref{phase_dagram}g). The helices in this phase form the hexagonal lattice ($n_c=6$) with two nearest like-chiral neighbours and four nearest opposite-chiral neighbours. The interaction phase for the like-chiral neighbours is the optimal value ($\Delta \Psi_{RR}=0$), while it is  non-optimal for the opposite-chiral pairs ($\Delta \Psi_{RL}=\pi/3$). 

At last, when $p$ further increases, the valley of the curve $d_{RR}(\Delta\Psi_{RR})$ becomes deeper and its minimum approaches that of $d_{RL}(\Delta\Psi_{RL})$ at $\Delta\Psi_{ij}=0$. 
In this case, the binary fluid first undergoes an entropy-driven chiral-separation (Figure~\ref{phase_dagram}h), and in each phase separated liquid, the self-rotating orientations of all helices are the same (see $\psi^{RO}_1$ in Figure~\ref{EOS}f).
This chiral-separated fluid state lacks the long-range position order, and the $\psi^{BO}_6$ of the system is low (Figure~\ref{EOS}f). Therefore, we refer this chiral-separated in-plane liquid crystal state as the separated-L (liquid) phase. In this phase, particles are best packed when each helix interacts with its like-chiral neighbours through the same zero interaction phase, and the system stays in this in-plane liquid crystal state before transforming to crystalline states at very high pressure as indicated by the increase of $\psi^{BO}_6$ in Figure~\ref{EOS}f. This chiral liquid crystal phase (also named as screw liquid crystal) has recently been explored extensively both experimentally\cite{Dogic2006PRL} and theoretically. \cite{cinacchi2016,Ferrarini_kolli2016,Ferrarini2014self}

For all these self-assembly processes, we find the density jump in the equation of state associated with the sudden change of the corresponding crystalline order (Figure~\ref{EOS}). This implies that the formations of these structures are all first-order phase transitions, which can be intuitively understood from the perspective of optimal packing. However, not all self-assembled structures here are the best packed structures of the corresponding systems at infinitely high pressure, and we also use MC annealing method obtaining the close-packed phase diagram of the system (Supplementary Figure~S1). Besides these order structures, we also find disordered amorphous (glass) states (Figure~\ref{phase_dagram}i). In this state, two chiral helical species mix randomly without any long-range order, and the crystal nucleation is not observed within our simulation time before the formation of amorphous solids.

\begin{figure*}[t]
\centering
  \includegraphics[height=12.0cm]{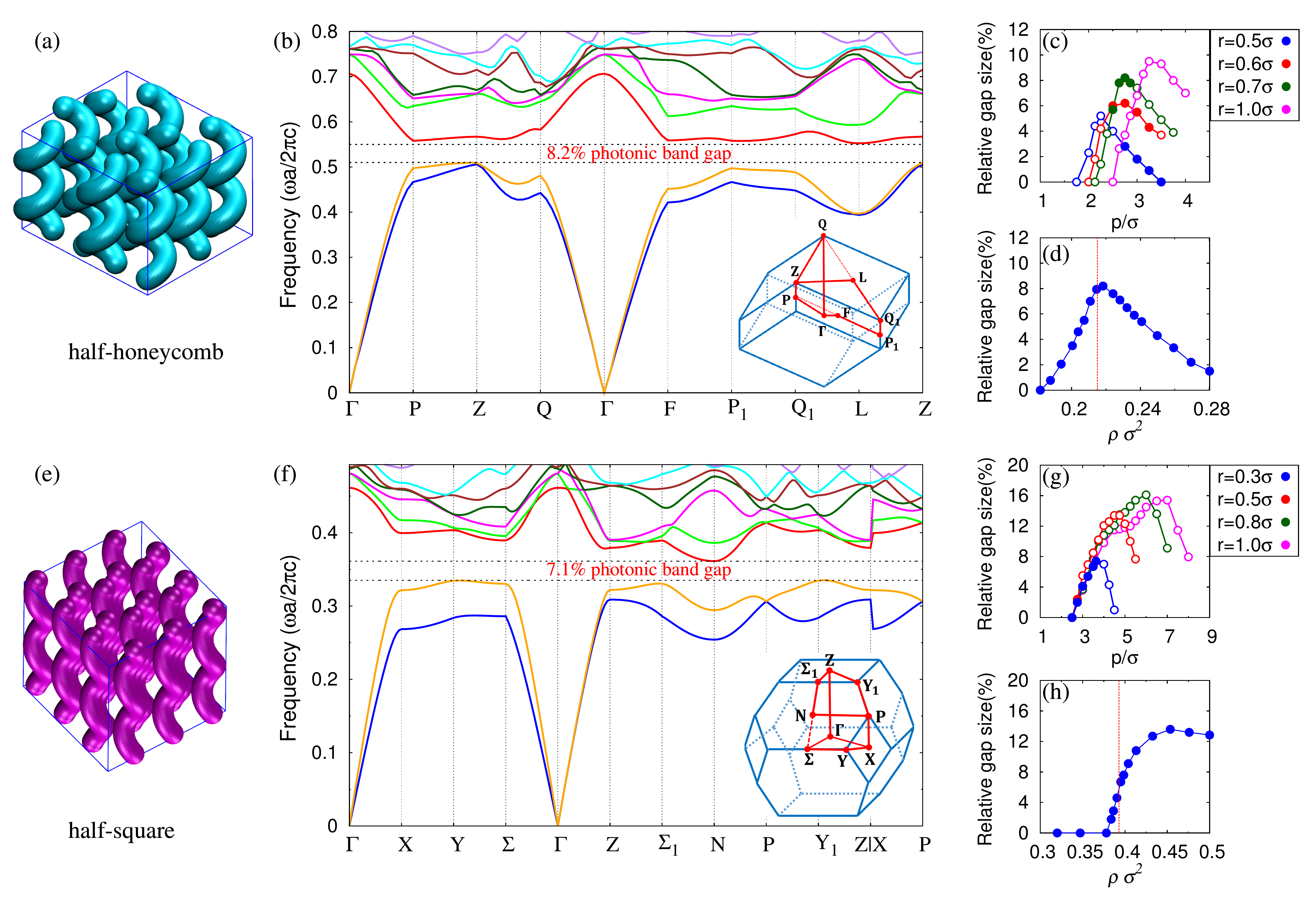}
  \caption{\label{3D_PBG} (a), (e): Close-packed half-honeycomb with $[r, p]=[0.7 \sigma , 2.75 \sigma ]$ and half-square crystal with $ [r, p] = [0.3\sigma, 3.5\sigma]$ and corresponding photonic band structures (b) (f).
Here $\omega$ and $c$ are the angular frequency and speed of the light with $a$ the lattice constant of the crystal in 2D. 
The high-symmetry points on the sweep paths in the band diagrams are marked in the first Brillouin zone of the corresponding crystal shown in the insets. (c),(g): Relative gap sizes as a function of $p$ with different $r$ for half-honeycomb and half-square chiral crystals, respectively, where the solid symbols represent the structures obtained in the self-assembly diagram Figure~\ref{phase_dagram}b. (d),(h): The relative gap sizes as functions of crystal density $\rho \sigma^2$ for half-honeycomb  and half-square chiral crystals with the same [$r$, $p$] as in (a) and (e), respectively. The red dashed lines indicate the close-packed densities.  All the figures are calculated at the dielectric contrast $\epsilon_r = 12$.}
\end{figure*}

{Lastly, we stress that in our simulations we neglect the tilt of helices to study the crystallization of helices racemate and local structures of chiral helices phase. However, this approximation cannot be applied to investigate the cholesterol phase of like-chiral helices at much larger length scales, in which the tilt of helices plays an important role.\cite{cinacchi2017,Frezza2014left,Dijkstra2014,dussi2015}}

\subsection*{3D PBGs in Self-Assembled Chiral Helical Crystals}

\begin{figure}
\centering
		\resizebox{80mm}{!}{\includegraphics[trim=0.0in 0.0in 0.0in 0.0in]{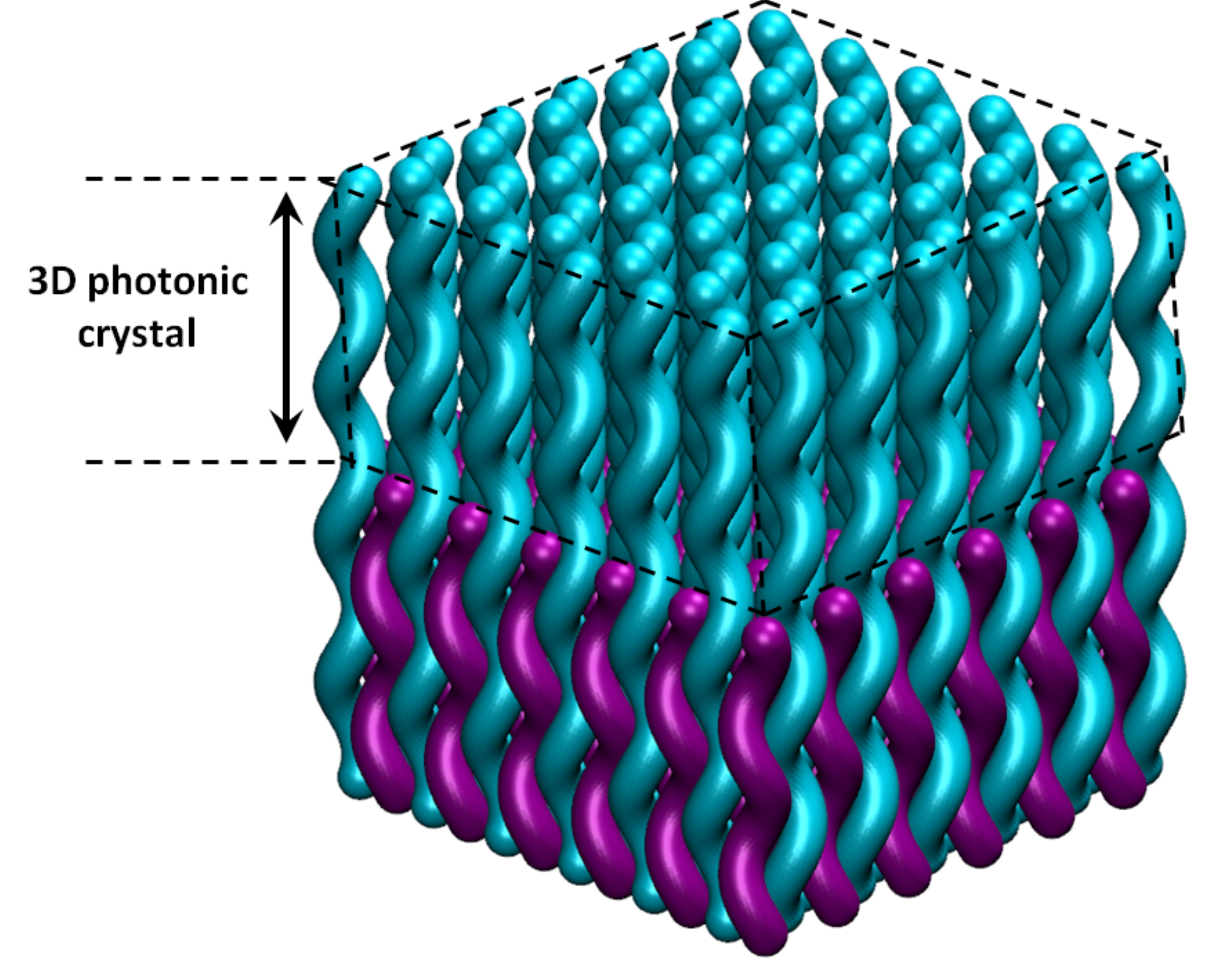} } 
\caption{\label{experiment} An experimental design to realize the 3D chiral photonic crystal based on the self-assembly of long and short colloidal helices with opposite handedness. { Due to the Brownian motion of colloidal helices, external fields, \textit{e.g.},  gravitational or electric field, are needed to help the sedimentation of short helices in order to obtain the chiral photonic crystal on the top.} }
\end{figure}

As mentioned in the previous section, both the self-assembled square and honeycomb helical crystals are essentially an interlock of two sets of chiral crystal with opposite handedness (see Figure~\ref{3D_PBG}a,e). After checking the photonic properties of all obtained crystals, we find complete 3D PBGs for these half-square and half-honeycomb helical crystals. In the half-square crystal, each helix has four $180^{\circ}$  phase-shifted nearest neighbours forming the symmetry of a body-center-tetragonal crystal. In the half-honeycomb crystals, helices have three types of phases ($0^{\circ}, \pm 120^{\circ} $). Helices of the same phases form hexagonal layers and  layers with different phases stack in \textbf{ABC} sequence along the axial direction (rhombohedral symmetry). The helices in most of the self-assembled chiral crystals are inter-contacted, and mechanically self-supported.  Experimentally, these half-square/honeycomb crystals can be obtained by dissolving one chiral species that is made up by different materials from their counterparts.\cite{vlasov2001chip,uday2017tunable} Another possible way that does not require the selective removal relies on the self-assembly of long and short colloidal helices in external field as proposed in the Figure~\ref{experiment}.

In Figure~\ref{3D_PBG}e,f, we show the corresponding band structures that have the largest PBGs for the half-honeycomb and half-square crystals at the dielectric contrast $\epsilon_r = \epsilon_{helix}/\epsilon_{air} =12$. {The structures we used in the PBG calculation are the closed packed crystals chosen from the corresponding self-assembly regions of Figure~\ref{phase_dagram}b}. For both crystals, the bandgap opens between the second and third bands with the relative gap size (gap/midgap ratio) of $7\sim 8\%$.
The relative gap sizes for the close-packed half-honeycomb and half-square crystals with various $p$ and $r$ are shown in Figure~\ref{3D_PBG}c,g. The structures which can be obtained from self-assembly of hard helices based on Figure~\ref{phase_dagram}b are shown as solid symbols otherwise open symbols. One can see that there is a  broad parameter space for half-honeycomb crystal having complete PBGs, while the one for half-square crystal is relatively narrow.
Nevertheless, the potential PBG size of the close-packed half-square crystal can be as large as 16\%. These crystals of helices with potential large PBGs can be realized by making the opposite-chiral helices oppositely charged to stabilize the binary square lattice.\cite{leunissen2005ionic} Introducing additional soft repulsions between helices, \textit{e.g.}, electrostatic repulsion or grafting  polymers on surface of helices, can also stabilize the square crystal by increasing the effective diameter of helices. 
Moreover, we plot the relative gap size as a function of  density for the half-honeycomb and half-square crystals in Figure~\ref{3D_PBG}d,h, in which the close-packed densities are shown as the vertical red dashed lines. We find that the PBG of the half-honeycomb crystal peaks at the close-packed density, while half-square crystals have larger PBG when overlaps between helices are allowed. If we further optimize the PBG size by changing $r$, $p$ and the overlap, the largest relative gap size can be 27\% (Supplementary Figure~S2). In experiments, increasing the overlap between colloidal particles can be realized by sintering the particles in the crystal.\cite{miguez1998control}

\begin{figure}[t]
\centering
  \includegraphics[height=6.0cm]{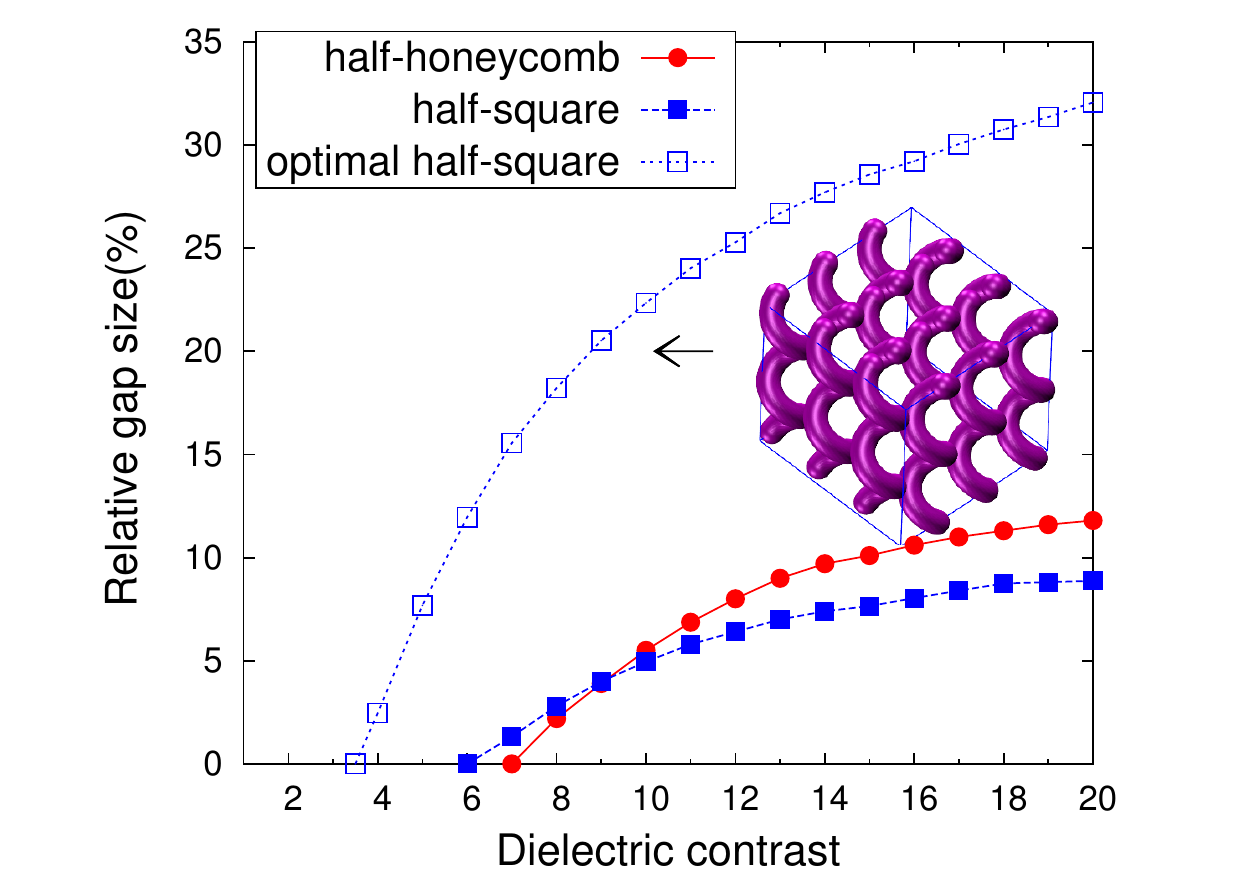}
  \caption{\label{dielectric}  Relative gap sizes as a function of dielectric contrast. The solid symbols show the relative gap sizes of the self-assembled half-honeycomb and half-square chiral crystals in Figure\ref{3D_PBG}, respectively, and the open symbols are for the optimal half-square chiral crystal of overlapped helices whose PBG opens at the lowest dielectric contrast ($\epsilon_r=3.5$). The optimal half-square chiral crystal  has the diamond symmetry ($p=\sqrt{2}a$) with $[r,p]=[0.7\sigma,3.88\sigma]$.}
\end{figure}

In Figure~\ref{dielectric}, we plot the relative gap size as a function of the dielectric contrast for the same crystals as in Figure~\ref{3D_PBG}a,e. We find that the miminal dielectric contrast required to have PBGs is about 7 and 6 for self-assembled half-honeycomb and half-square crystals, respectively. In the same figure, we also plot the data for the optimal half-square crystal of overlapped helices with $[r,p]=[0.7\sigma,3.88\sigma]$. The PBG of this structure closes at the lowest dielectric contrast of $3.5$, and this low-dielectric performance is even better than the optimized diamond crystals of dielectric spheres.\cite{joannopoulos2011p,hynninen2007self} Actually this optimized half-square structure is also of the diamond symmetry.\cite{maldovan2004d}
This offers  possiblities to fabricate photonic crystals with PBGs in the visible region by using low-refractive-index materials, such as titanium dioxide ($n_r=2.6$), moissanite ($n_r=2.7$), zinc sulphide ($n_r=2.36$), gallium nitride ($n_r=2.4$) and zinc oxide($n_r=2.4$) \textit{etc.}, whose dielectric contrast in air $\epsilon_r = n_r^2$.\cite{rii} Other common materials like alumina, mica and some commercial glass can also be used to fabricate photonic crystals in the microwave region with possible applications in all-dielectric antenna.\cite{kesler1996antenna} Although this work mainly focuses on the ordinary dielectric materials, more complicated optic phenomena are expected in  systems of metallodielectric helices,\cite{metal_helix2014,esposito2014n,optic2010PRL,optic2011PRL,
science2009goldhelix} and core-shell nano-helices.\cite{kosters2017core}

\subsection*{Polarization PBGs in Self-Assembled Helical Crystals}

\begin{figure}[t]
\centering
  \includegraphics[height=8.0cm]{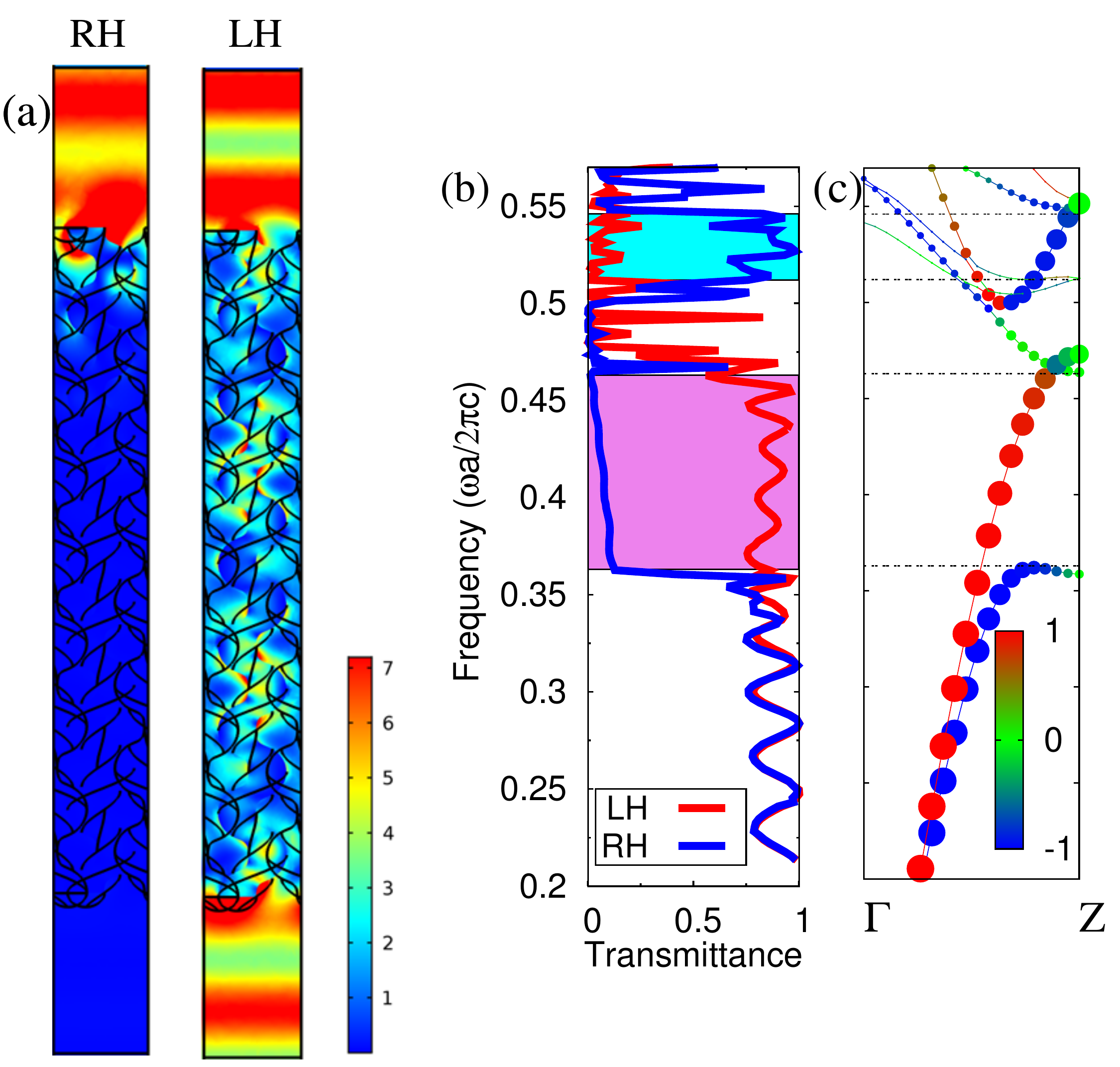}
  \caption{\label{polarization}  (a) { Side view of optical simulation} of normal incident circular polarized left-handed (LH) and right-handed (RH) lights with frequency $\omega a/2\pi c =0.45$ upon {a monolayer of} the close-packed RH half-square crystal $[r,p]=[0.4\sigma,2.25\sigma]$. {The source incident light propagates from top to down}. The color represents the intensity of magnetic field (A/m). {The weak magnetic field on the bottom side of crystal layer for the  RH light case on the left side indicates the forbidden transmission, which is in contrast with the LH light case on the right side.} (b) Transmission  spectra  of LH (red) and RH (blue) circularly polarized lights. (c) the band structure along $\Gamma-\mathrm{Z}$ direction. The color and size of the symbols indicate the CD index $\alpha$ and the coupling index $\beta$, respectively, with the largest symbol of $\beta=0.9$. The calculation is performed at the dielectric contrast $\epsilon_r = 12$.}
\end{figure}

{Sparse} helices arranged in square and hexagonal lattices with the zero phase-shift were shown to have large circular polarization photonic bandgaps in the direction of the helical axis.\cite{lee2005pol,kao2015d} When incident lights enter from this direction, the helical array can inhibit the propagation of one specific circularly polarized mode but allow the transmission of its oppositely handed counterpart. However, {few} self-assembled structures {were reported to have} such polarization bandgap. 
In Figure~\ref{polarization}a,b, we show the simulation snapshot and transmission spectra of circularly polarized lights through the self-assembled closed-packed half-square crystal of right-handed helices with $[r,p]=[0.4 \sigma, 2.25\sigma]$ at the normal incident. We find dual polarization PBGs open for right-handed (RH) and left-handed (LH) polarized lights separately at different frequency regions, as marked by cyan and magenta bands in the transmssion spectra in Figure~\ref{polarization}h. The relative sizes are about 24\% and 6.5\% for the low and high frequency gaps, respectively. In Figure~\ref{polarization}a, we also show the distribution of magnetic field for LH and RH light with frequency $\omega a/2\pi c =0.45$, where a strong chiral selectivity can be found for the light to pass the crystal. 

To understand the physics of these dual polarization gaps, we perform the circular polarization analysis  by calculating the circular dichroism (CD) index $\alpha$ and the coupling index $\beta$  of the eigenmodes of this photonic crystal.  $\alpha$ and  $\beta$ are defined as \cite{lee2005pol,kao2015d}
\begin{eqnarray}
\alpha &=& \rm{sgn}(\textbf{q} \cdot \nabla_\textbf{k} \omega ) \frac{\eta_R - \eta_L}{\eta_R + \eta_L}, \\
\beta &=& \eta_{R} + \eta_{L},
\end{eqnarray}
with 
\begin{eqnarray}
\eta_{R/L}= \frac{\left \vert \int \int (\textbf{e}_{x} \mp i\textbf{e}_{y}) \cdot \mathbf{H}(x,y,z_0)dxdy \right \vert^2}{\int \int \left \vert (\textbf{e}_{x}\mp i\textbf{e}_{y})\right \vert^2 dxdy  \int \int \left \vert \mathbf{H}(x,y,z_0)\right \vert^2 dxdy}
\end{eqnarray}
where $\mathbf{H}(x,y,z_0)$ is the magnetic field on the largest cut plane of the crystal unit cell at $z_0$. The CD index $\alpha$ ranged $[-1,1]$ reflects the preference that the Bloch modes coupled to RH (-1) or LH (+1) light, while $\beta$ ranged [0,1] measures how strong this coupling is. We plot the band structures in Figure~\ref{polarization}c, where we use the color and size of the symbols to represent the value of $\alpha$ and $\beta$, respectively. We find that the polarization bandgap for RH lights at low frequency region is a circular dichroic bandgap open between the first and second bands, in which only strong coupled LH eigenmodes exist, while the polarization bandgap for LH lights at relatively higher frequency region is a low-coupling stop band due to the low-coupling of LH plane wave with the Bloch modes rather than the exclusive RH eigenmodes.\cite{saba2011} In fact, all self-assembled half-square crystals at closed-packed density exhibit polarization bandgap between the first and second bands based on the eigenmodes analysis, indicating robust circular dichroism effects in this self-assembled structure. This effect can be potentially utilized to fabricate chiral beamsplitters as the logic gate in photonic quantum technologies where `qubits' are represented by circularly polarized states of photons.\cite{o2009photonic,lodahl2017}

{\section*{conclusions}}
In conclusion, with the help of a {simplified} simulation method, we study a bottom-up self-assembly strategy for large-scale fabrication of chiral photonic crystals. This strategy is based on the entropy-driven co-crystallization of oppositely chiral colloidal hard helices.  Depending on the radius and pitch of helices, the racemic helices fluid spontaneously nucleates into binary crystals with various symmetries at high density. Two self-assembled chiral structures are shown to have large complete 3D PBGs, and dual polarization PBGs are also observed in one of these self-assembled structures. The parameter space in which the self-assembled crystals have PBGs is broad, which gives large freedom for the experimental realization. {In our system of smectic helices monolayer, the alignment of helices effectively reduces the 
 freedom degrees of particles in the system}.  The nucleation barriers, therefore, are relatively low, and the spontaneous {crystallization} can be observed  in direct computer simulations. This means that no additional surface patterning\cite{hynninen2007self} or pre-designed interaction\cite{ducrot2017colloidal,avvisati2017,patta2017novel} is required to induce the crystallization. These special conditions have been shown crucial in the directed self-assembly of diamond structure of spherical colloids due to the small free energy difference between the desired crystal and other analogues.\cite{hynninen2007self,ducrot2017colloidal}  Lastly, although we focus on the self-assembly of hard helices, the chiral interaction we find here is generic and similar self-assembly mechanism is expected to exist in other helical or spiral shaped particles systems, as well as between long semi-flexible helical chains, which have promise in building up mechanic-responsive photonic crystal fibre sensor\cite{xi2013measuring,zhang2017highly} and helical optical fibers with topological functionalities.\cite{alexeyev2013spin,kartashov2013dynamics,
alexeyev2016localized,petrovic2017rotating,lu2016topological}

\ \\

\section*{methods}
\subparagraph*{Proof of Equation~(1) and Equation~(2) }
Assuming the axis of the helix is along the $z$ direction, the central line of the reference helix $i$ with arbitrary handeness ($i=R,L$), which passes the central points of cross-sections of the helix in the xy plane, can be described by the parametric equations with the parameter $t$
\begin{eqnarray} 
u^x_i(t)&=& x_i + r \cdot \cos(C_i t+\phi_i), \label{helix_eqt1} \\
u^y_i(t)&=& y_i + r \cdot \sin(C_i t+\phi_i) ,\label{helix_eqt2} \\
u^z_i(t)&=& z_i + \frac{p}{2\pi} \cdot t, \label{helix_eqt3}
\end{eqnarray}
where $\textbf{u}_i=[u^{x}_i,u^{y}_i,u^{z}_i]$ is the coordinate of the point on the central line of helix $i$, and $x_i,\ y_i,\ z_i$ are the coordinate  of {helical axis}. $r$ and $p$ are the helical radius and pitch, respectively. $C_i$ is the sign of chirality ($C_R=1, C_L=-1$) and $\phi_i$ is the self-rotating angle of helix $i$. {By setting the axis at the origin point, \textit{i.e.}, $x_i=y_i=0$}, the parametric Equation (\ref{helix_eqt1}-\ref{helix_eqt3}) can be simplified to
\begin{eqnarray}
u^x_i &=&r \cdot \cos \left[  \frac{2 C_i \pi }{p} (u^z_i-z_i) +\phi_i \right],  \\
u^y_i &=&r \cdot \sin \left[  \frac{2 C_i \pi }{p} (u^z_i-z_i) +\phi_i \right].
\end{eqnarray}
The screw axis symmetry of helix, \textit{i.e.}, a $ \Delta \phi_i$ self-rotating is equivalent to the $ \frac{C_i \Delta\phi_i}{2  \pi } p$ elevation in $z$ direction, can be written as:
\begin{eqnarray}
\textbf{u}_i(z_i,\phi_i+ \Delta\phi_i)=\textbf{u}_i(z_i+\frac{C_i \Delta\phi_i}{2  \pi } p,\phi_i).
\end{eqnarray}

When another helix $j$ is approaching helix $i$, the contact distance $d_{ij}$ between the axes of two helices can be generally written as a periodic function of five  independent variables: $\phi_i$, $\phi_j$, $\theta$, $z_i$, $z_j$, where $\theta$ is the angle of position of helix $j$ with respect to helix $i$. $d_{ij}$ does not change when two helices have the same displacement in the $z$ direction, and this translational symmetry reduces the five independent variable to four:
\begin{eqnarray}
&& d_{ij}(\phi_{i},\ \phi_{j},\ \theta, z_{i},z_{j} )=d_{ij}(\phi_{i},\ \phi_{j},\ \theta, z_{j}-z_{i} ).
\end{eqnarray}
Further considering the aforementioned screw axis symmetry reduces the four variables  to two,
\begin{eqnarray}
&& d_{ij}(\phi_{i},\ \phi_{j},\ \theta, z_{j}-z_{i} )= \nonumber 
\\ 
&& d_{ij}(\theta,\ C_j\phi_j + 2\pi z_{j}/p - C_i\phi_i - 2\pi z_{i}/p ).
\end{eqnarray}

\begin{figure}
\centering
		\resizebox{60mm}{!}{\includegraphics[trim=0.0in 0.0in 0.0in 0.0in]{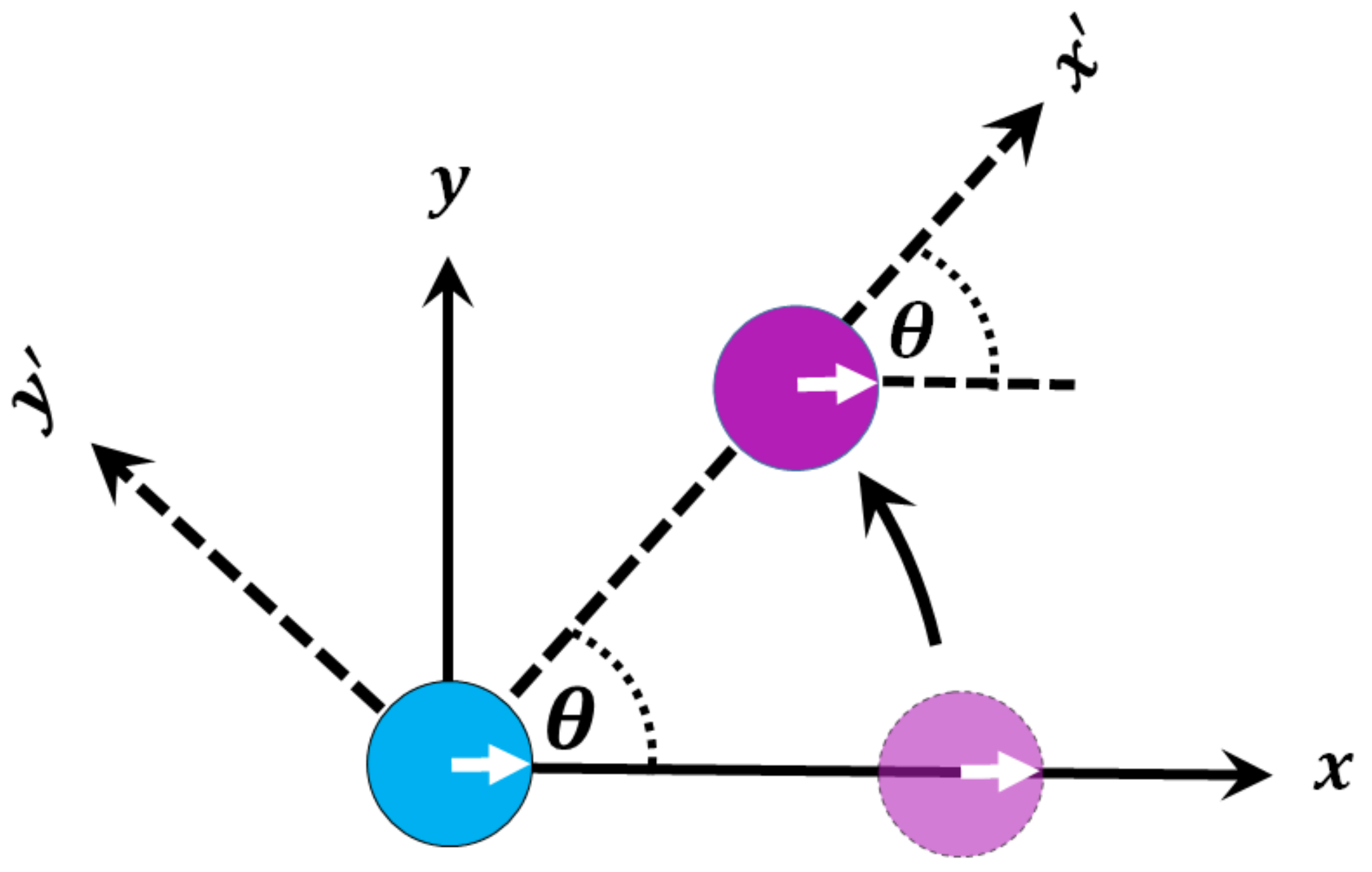} } 
\caption{\label{method} The effect of angle of position $\theta$ on the pairwise interaction of two helices (blue: reference helix $i$, magenta: approaching helix $j$). The self-rotating angles of the helices are fixed during the change of angle of position $\theta$.}
\end{figure}

To prove that the remaining two variables, in fact, are linear dependent too, we study the contact of two helices in a reference coordinate system $x-y$ where helix $j$ approaches the reference helix $i$ from the $x$ direction which is along the vector that connects the axes of two helices, as shown in Figure~\ref{method}. We fix $\phi_i$, $\phi_j$ in this reference coordinate and change the angle of position of helix $j$ with respect to helix $i$ from 0 to $\theta$. The contact distance between two helices can be recalculated in a new reference coordinate system $x'-y'$ represented by dashed arrows in Figure~\ref{method}. In this new coordinate system, the self-rotating angles for both helices decrease $\theta$ compared with that in the old coordinate system:\begin{eqnarray}
\phi'_i &=& \phi_i-\theta, \\
\phi'_j &=& \phi_j-\theta.
\end{eqnarray}
Since the contact distance is independent of the choice of the reference coordinate system, we can have
\begin{eqnarray}
&& d_{ij}(\theta,\ C_j\phi_j + 2\pi z_{j}/p - C_i\phi_i - 2\pi z_{i}/p ) =  \nonumber
\\
&& d_{ij}[0,\ C_j(\phi_j-\theta) + 2\pi z_{j}/p - C_i(\phi_i-\theta) - 2\pi z_{i}/p ]. \nonumber \\
\end{eqnarray}
This indicates that $d_{ij}$ can be written as a function of a single variable which is a linear combination of $\phi_i,\ \phi_j,\ \theta,\ z_i,\ z_j$, 
\begin{eqnarray}
&d_{ij}(\phi_{i},\phi_{j},\theta, z_{i},z_{j} )= d_{ij}\left (\Delta \Psi_{ij}\right), \\
&\Delta \Psi_{ij} = C_j\phi_{j}-C_i\phi_{i}+(C_j-C_i)\theta + 2 \pi (z_{j}-z_{i})/p \ \ \ \ \ \ \ 
\end{eqnarray}

\subparagraph*{Monte Carlo simulations}
In the computation of contact distance $d_{RR/RL}$,  we fix  {the reference} helix (both position and self-rotation) and move the other helix with the same/opposite chirality towards the reference one to search for $d_{RR/RL}$ using the bisection method. Each helix is composed by 1000 spheres per pitch to make helix smooth enough to mimic the ideal helix. We tabulate the interaction phase $\Delta \Psi_{ij}$ from 0 to $2\pi$ with the step size $\pi/360$, and calculate the corresponding $d_{RR/RL}$, which we use with the linear interpolation in the MC simulations to check overlap between helices. {In the MC simulation of the helices' self-assembly, we employ periodic condition in $xyz$ direction and the Isobaric-Isothermal (NPT) ensemble to simulate binary systems of $N=400 \sim 1024$ hard helices. The box lengths are allowed to change independently in $x,y$ direction. During the simulation, translational trial moves in $xy$ plane, rotational trial moves around helical axis, shift trial moves in the axial directions, as well as the box length change in $x, y$ directions are randomly selected. These trial movements are accepted based on the classical Metropolis algorithm.\cite{frenkel2001understanding} A self-adapted method is used to find the optimal step sizes for each kind of trial movement during the equilibiration of the system. Then the step sizes are fixed in the sampling process. The cell-list technique\cite{frenkel2001understanding} is also used to accelerate the simulation.} In our simulation, the initial configuration of the system is a disordered racemic fluids. We then slowly increase the pressure of the system and identify the first ordered phase that nucleates. 
{To observe the nucleation and calculate the EOS, we perform MC simulations up to order of $10^8$ MC sweeps}. The $n$-fold bond orientational order parameter is defined as $\psi^{BO}_n=\left \langle  \left| \frac{1}{N}\sum^{N}_{i=1} \sum^{n}_{j=1} \exp(i n \theta_{ij} )  \right|  \right\rangle$),
where $\theta_{ij}$ is the angle between $\mathbf{r}_{i} - \mathbf{r}_{j}$ and the $x$-axis with 
{ $\mathbf{r}_i=[x_i,y_i]$ and $\mathbf{r}_{j}=[x_j,y_j]$ (see Equation~(\ref{helix_eqt1}),(\ref{helix_eqt2})) the position of helix $i$ and one of its first $n$ nearest neighbour $j$, respectively. The self-rotating orientational order parameter is defined as $\psi^{RO}_1=\left \langle  \left| \frac{2}{N}\sum^{N/2}_{i=1} \exp(i \phi_i )  \right|  \right\rangle$, where the summation runs over helices of the same chirality. }

\subparagraph*{Optical simulations}
{ The photonic band structures are calculated by solving the eigenvalue problem of Maxwell's equations for electromagnetic fields.\cite{joannopoulos2011p} Since we focus on ordinary dielectric materials, we ignore frequency dependence of dielectric constant $\epsilon$ and magnetic permeability $\mu$ and  assume both $\epsilon$ and $\mu$  purely real and positive. The Maxwell's equations thus can be simplified as an equation of magnetic field $\mathbf{H}(\mathbf{r,t})= \mathbf{H}(\mathbf{r})e^{i\omega t}$ in the frequency domain as\cite{joannopoulos2011p}
\begin{eqnarray} \label{Max_H}
\nabla\times\left[\frac{1}{\epsilon(\mathbf{r})}\nabla\times \mathbf{H}(\mathbf{r})\right]=\left( \frac{\omega}{c} \right)^2 \mathbf{H}(\mathbf{r}),
\end{eqnarray}
with light speed $c=\sqrt{\epsilon \mu}$. For periodic structures,  according to the Bloch's theorem, the eigenvector of Equation~(\ref{Max_H}) for specific wave vector $\mathbf{k}$ is proportional to periodic Bloch envelope $\mathbf{u_k}$ as $\mathbf{H_k}(\mathbf{r}) = e^{i\mathbf{k}\cdot \mathbf{r}}\mathbf{u_k}(\mathbf{r})$  with transversality constraint $(i\mathbf{k}+\nabla) \cdot \mathbf{u_k}=0$. Equation~(\ref{Max_H}) thus can be rewritten  as\cite{joannopoulos2011p}
\begin{eqnarray} \label{Max_Bloch}
\left[(i\mathbf{k}+\nabla)\times \frac{1}{\epsilon(\mathbf{r})}(i\mathbf{k}+\nabla)\times \right] \mathbf{u_k}(\mathbf{r})= \frac{\omega(\mathbf{k})^2}{c^2} \mathbf{u_k}(\mathbf{r}). \nonumber \\
\end{eqnarray}
The solution of Equation~(\ref{Max_Bloch}) yields $\omega(\mathbf{k})$ as a function of $\mathbf{k}$ which is the photonic band structure of the crystal.  The sweep paths and high symmetry points in the first Brillouin zone are chosen based on Ref.~\onlinecite{setyawan2010high}. Since $\mathbf{u_k}$ is a periodic function, one only needs to solve the partial differential equation of Equation~(\ref{Max_Bloch}) in a  single discretized crystal unit cell with iterative method.}

{The transmission spectra in Figure~{\ref{polarization}}b are calculated directly based on Equation~(\ref{Max_H}) with the Floquet periodicity in $xy$ directions. As shown in Figure~{\ref{polarization}}a, the thickness of photonic crystal layer is 8 unit cells. The perfectly matched layers (PML)\cite{berenger1994perfectly} on the top and bottom of the simulation box are used to absorb the excess reflected and transmitted light wave. The source port 1 for normal incident polarized lights is located between the top PML and crystal layer. The listener (output) ports 2, 3 for transmitted LH and RH polarized lights are put between the crystal layer and bottom PML. The transmittance for LH and RH polarized light is calculated as
\begin{eqnarray} \label{Port}
T_{L}=|S_{21}|^2,\ \ \ \ \ \ \ \ T_{R}=|S_{31}|^2,
\end{eqnarray}
with the scattering parameters $S_{ij}$ connecting the source port $j$ and output port $i$. For all the simulations, finite elements method (FEM)\cite{jin2015finite} is used as the discretization scheme and we ensure that all results are converged with respect to the discretization degree.}

\

{\bf Supporting Information} The Supporting Information is available on the website at DOI: XXXXXXXXXX.

{
Figures S1-S2, close-packed phase diagram of helices racemate  and photonic band structures for bandgap-optimized half-square crystal. }

\begin{acknowledgments}
The authors thank Prof. Huanyang Chen, Prof. Yidong Chong, Dr. Zhiguo Yang, Dr. Fei Gao and Shampy Mansha for helpful discussions on optical simulations.
This work is supported by the National Natural Science Foundation of China (No.~91427302, 11474155, and 11774147), Nanyang Technological University Start-Up Grant (NTU-SUG: M4081781.120), the Academic Research Fund Tier 1 from Singapore Ministry of Education (M4011616.120 and M4011873.120), and the Advanced Manufacturing and Engineering Young Individual Research Grant (A1784C0018) by the Science and Engineering Research Council of Agency for Science, Technology and Research Singapore.
\end{acknowledgments}

\bibliographystyle{ACS}
\bibliography{helix,optical}

\begin{figure*}
\centering
		\resizebox{150mm}{!}{\includegraphics[trim=0.0in 0.0in 0.0in 0.0in]{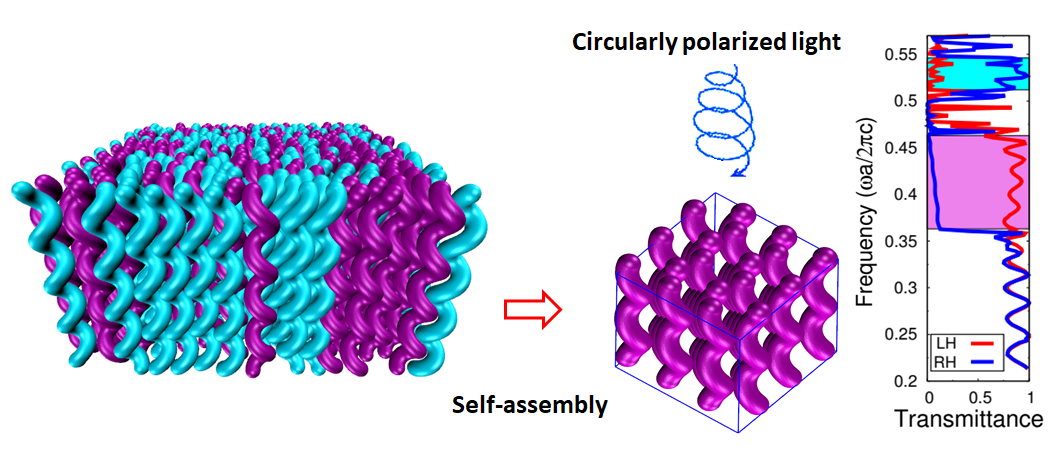} } 
\caption*{TOC: Liquid crystal monolayer composed of opposite-handed helices  self-assembles into chiral photonic crystal with 3D full photonic bandgaps and polarization selectivity.  }
\end{figure*}

\end{document}